\documentstyle[12pt]{article}
\begin{document}
\begin{flushright}
arXiv: 0802.4129 [hep-th]\\ BHU-PHYS-CAS/Preprint
\end{flushright}
\vskip 2.5cm
\begin{center}
{\bf \Large {Free Abelian 2-form gauge theory: BRST approach}}

\vskip 3.5cm

{\bf R. P. Malik} \\
{\it Centre of Advanced Studies, Physics Department,}\\ {\it
Banaras Hindu University, Varanasi-221 005, U. P., India}\\ 
{\small {\bf E-mails: malik@bhu.ac.in ; rudra.prakash@hotmail.com}}

\vskip 1.5cm
\end{center}

\noindent {\bf Abstract}: We discuss various symmetry properties
of the Lagrangian density of a four (3 + 1)-dimensional (4D) free Abelian
2-form gauge theory within the framework of Becchi-Rouet-Stora-Tyutin (BRST)
formalism. The present free Abelian gauge theory is endowed with a Curci-Ferrari type
condition which happens to be a key signature of the 4D non-Abelian 1-form gauge theory.
In fact, it is due to the above condition that the nilpotent BRST and anti-BRST symmetries
of our present theory are found to be {\it absolutely} anticommuting in nature. 
For the present 2-form theory, we discuss the BRST, anti-BRST, ghost 
and discrete symmetry properties of the Lagrangian densities and derive 
the corresponding conserved charges. The algebraic structure, obeyed by
the above conserved charges, is deduced and the
constraint analysis is performed with the help of physicality criteria
where the conserved and nilpotent (anti-)BRST charges play completely {\it independent}
roles. These physicality conditions lead to the derivation of the above Curci-Ferrari
type restriction, within the framework of BRST formalism, from the constraint analysis.\\

\baselineskip=16pt

\vskip .7cm


\vskip 1.7cm

\newpage

\noindent
{\bf 1 Introduction}\\

\noindent 
The Becchi-Rouet-Stora-Tyutin (BRST) formalism is one of the key theoretical
tools that plays a pivotal role in dealing with many mathematical and physical
facets of the gauge theories, topological
field theories, reparametrization invariant theories, string theories, etc.
In the realm of the 4D non-Abelian 1-form gauge theories (that lay the cornerstones 
for the whole edifice of the standard model of high energy physics), 
the BRST approach provides the proof of the unitarity of the quantum gauge theory. To
be precise, in this approach, the local gauge 
symmetries of the original classical theory (that are generated by 
the first-class constraints [1,2] of the theory) are traded with the ``quantum''
BRST symmetry transformations which turn out to be nilpotent of order two 
(see, e.g. [3-6]).

A well-defined set of nilpotent BRST symmetry transformations plays a crucial role in the
derivation of the Slavnov-Taylor (ST) identities in the realm of the 4D non-Abelian
1-form gauge theories [7,8]. In some sense, the Ward-Takahashi identities for
QED are the limiting case of the ST identities and the former identities 
also owe their origin to the BRST-type symmetries. These identities play decisive roles
in proving the renormalizability of the above 4D (non-)Abelian 1-form gauge theories.
It would be an interesting endeavor to explore the ramification of the 
application of the BRST approach
to the 2-form $B^{(2)} = [(dx^\mu \wedge dx^\nu)/2!] B_{\mu\nu}$)     
gauge theories. One of the central themes of our present investigation is to accomplish
this goal in a clear and cogent fashion. In particular, it would be interesting to find
out, within the framework of BRST formalism,
some new features in the domain of 2-form gauge theories that are different from the
1-form gauge theories where the BRST approach has been so successful. We do find
some new results in our present endeavor.

In the recent past, the study of antisymmetric ($B_{\mu\nu} = - B_{\nu\mu}$) tensor 
gauge potential $B_{\mu\nu}$ has become quite popular because of its 
relevance in the context 
of the modern developments in string theories and related extended objects. 
This potential appears in the excited states of the quantized (super)strings,
multiplets of the supergravity theories [9] and plays a crucial role in providing
the noncommutativity in the string theory [10]. The 4D gauge potential $B_{\mu\nu}$ leads to
a dual description of the massless scalar field [11-13] and generates mass for the 
Abelian 1-form ($A^{(1)} = dx^\mu A_\mu$) gauge potential through the celebrated topological
(i.e. $B \wedge F$) term. In the latter case, the Abelian U(1) gauge invariance and mass of the gauge field
co-exist together without any recourse to the Higgs mechanism. The 2-form gauge theory
has been studied from various points of view including the constraint Hamiltonian analysis [14,15] 
and BRST formulation [16-18].

The 2-form gauge theories are endowed with a rich mathematical structure. For instance,
the 4D Abelian 2-form gauge theory, in our earlier works [18,19], has been shown to
provide a cute field theoretical model for the Hodge theory where all the de Rham cohomological
operators of the differential geometry find their counterparts in the language of the conserved
charges that generate a set of well-defined symmetry transformations for a specific Lagrangian
density of the theory. Furthermore, this theory has been shown to correspond to a quasi-topological
field theory where the topological invariants have been shown to have proper recursion relations
[20]. The gauge symmetry transformations and corresponding BRST symmetry transformations for this
theory have also been discussed in the framework of Wigner's little group [21]. In all the above works
[16-21], however, the nilpotent (anti-)BRST symmetry transformations have been found to be anticommuting
{\it only} up to the Abelian U(1) vector gauge transformations.

In our very recent work [22], we have applied the superfield approach to 4D free Abelian
2-form gauge theory. The innate strength of this approach is to provide the BRST and anti-BRST
symmetry transformations that are always found to be nilpotent of order two and {\it absolutely} 
anticommuting in nature. The outcome of this work has led to the derivation of a relationship 
(cf. (3.4) below) that is just the analogue
of the Curci-Ferrari restriction of the 4D non-Abelian 1-form gauge theory [23]. This is
a completely new result because the application of the other superfield approaches [24-26]
to the 4D Abelian 2-form gauge theories
do {\it not} lead to such kind of relationship. In one of our very recent works [27], this
relationship has been taken into account in the Lagrangian density itself (cf. (3.1) and (4.1)
below) and it has been claimed that this type of relationship would always appear in the context
of the higher p-form ($p \geq 2$) {\it Abelian} gauge theories because of its intimate connection with the 
concept of geometrical objects called gerbes (see, e.g. [27]).

The purpose of our present investigation is to begin with the Lagrangian densities of
our earlier work [27] and find out the conserved charges corresponding to the off-shell nilpotent
and absolutely anticommuting (anti-)BRST symmetries of the theory. The physicality
conditions, with these conserved and nilpotent (anti-)BRST charges, lead to the derivation
of the Curci-Ferrari type of restriction from the constraint analysis. This is a new
way to derive this condition that has been deduced due to the application of the superfield
approach to BRST formalism in [26]. It turns out that the BRST and anti-BRST charges play
completely {\it independent} roles in this derivation. Thus, for our present gauge theory, the anti-BRST
charge is not merely a decorative piece in the theory. Rather, it has its own real identity. 
Furthermore, we
demonstrate that there is existence of a set of discrete symmetry transformations 
in the theory which plays a key role in providing us clues on how to go from BRST symmetry 
transformations to the anti-BRST transformations and vice-versa. We also find
out the continuous ghost symmetries of the theory and derive the BRST algebra. In the proof
of this algebra, once again, the Curci-Ferrari type restriction plays a very
decisive role. The cohomological aspects of this algebra are discussed briefly and
some of the issues, that are not resolved in our present work, are pointed out.

The following central factors have contributed to our main motivation in pursuing the
present investigation. First and foremost, in our earlier endeavors [27,22], we
have only mentioned the anticommuting (anti-)BRST symmetry transformations but
have not computed the generators of these transformations. Thus, it is essential
to compute the (anti-)BRST charges and, with their helps,
discuss the constraint structure of the theory. Second,
it has been challenging to derive the Curci-Ferrari type restriction, obtained
in [27,22], within the framework of the BRST formalism itself. In our present
work, we have derived the same from the physicality criteria. Third, our
present investigation is our first modest step towards our main goal of applying the 
BRST formalism to the 4D and 6D {\it non-Abelian} 2-form gauge theories that
are relevant in the context of string theories. Finally, our understanding of
the Lagrangian densities (cf. (3.1),(4.1)) and their symmetries might
help us in gaining some key insights that would enable us to derive
the dual(co)-BRST and anti-co-BRST symmetry transformations. These symmetries
would be sufficient, in our future attempts, to prove that the 4D Abelian 2-form
gauge theory is a precise model for the Hodge theory.

The contents of our present investigation are organized as follows. 

In Sec. 2,
we recapitulate the bare essential of the local, covariant and continuous 
gauge symmetries that are generated by the first-class constraints of the theory.

Section 3 is devoted
to the resolution of some the issues raised at the fag end of Sec. 2 within
the framework of the BRST formalism. The explicit expression for the BRST charge 
is derived here and the role of the constraints in the consistent
quantization of the theory is discussed with the help of 
the physicality condition with this charge. 

Section 4 deals with the anti-BRST
symmetry transformation, corresponding conserved charge and constraint
analysis that follow from the physicality condition with this charge. 
The independent identity of this charge is emphasized in the context 
of constraint analysis.

In Sec. 5, we derive the ghost charge and lay stress on the importance
of the discrete symmetries that are present in the theory. 

Our Sec. 6 focuses briefly
on the cohomological aspects of the BRST algebra and the derivation of the latter
in terms of the conserved charges of the theory. 

Finally, we make some concluding remarks in 
Sec. 7 and point out a few future directions for further investigations
in the realm of the 2-form gauge theories.

In the Appendices A and B, we derive the off-shell and on-shell nilpotent BRST 
as well as anti-BRST 
transformations for the appropriate Lagrangian densities of the 4D Abelian 2-form gauge
theory that do {\it not} respect, {\it together} with the above symmetries,
any kind of anticommuting anti-BRST as well as BRST symmetry
transformations, respectively.\\

\noindent
{\bf 2 Preliminary: gauge symmetry transformations}\\

\noindent
We begin with the well-known Kalb-Ramond Lagrangian density (${\cal L}^{(0)}$) for the 
4D\footnote{We follow here the conventions and notations
such that the 4D Minkowskian spacetime manifold is endowed with a flat metric with the signatures
(+1, -1, -1, -1) so that $A \cdot B = A_\mu B^\mu \equiv A_0 B_0 - A_i B_i$ is the dot product between two non-null 4-vectors $A_\mu$ and $B_\mu$. 
Here the Greek indices $\mu, \nu, \kappa..... = 0, 1, 2, 3$ correspond to the
spacetime directions and Latin indices $i, j, k....= 1, 2, 3$ stand for the space directions
only. We adopt here the differentiation convention $(\partial B_{\mu\nu} /\partial B_{\rho\sigma})
= \frac{1}{2!}(\delta_{\mu\rho} \delta_{\nu\sigma} - \delta_{\mu\sigma} \delta_{\nu\rho}),
(\partial H_{\mu\nu\kappa}/ \partial H_{\rho\sigma\eta}) = \frac{1}{3!} [\delta_{\mu\rho}
(\delta_{\nu\sigma} \delta_{\kappa\eta} - \delta_{\nu\eta} \delta_{\kappa\sigma})
+ \delta_{\mu\sigma} (\delta_{\nu\eta} \delta_{\kappa\rho} - \delta_{\nu\rho} \delta_{\kappa\eta})
+ \delta_{\mu\eta} (\delta_{\nu\rho} \delta_{\kappa\sigma} - \delta_{\nu\sigma} \delta_{\kappa\rho}),
$ etc., for the antisymmetric fields.}
antisymmetric ($B_{\mu\nu} = - B_{\nu\mu}$)
tensor gauge field $B_{\mu\nu}$, as [28,11-13]
$$
\begin{array}{lcl}
{\cal L}^{(0)} = {\displaystyle \frac{1}{6}} H^{\mu\nu\kappa} H_{\mu\nu\kappa}, \qquad
H_{\mu\nu\kappa} = \partial_\mu B_{\nu\kappa}
+ \partial_\nu B_{\kappa\mu} + \partial_\kappa B_{\mu\nu},
\end{array} \eqno(2.1)
$$
where the totally antisymmetric ($H_{\mu\nu\kappa} = - H_{\nu\mu\kappa} = - H_{\kappa\nu\mu}$, etc.)
curvature tensor $H_{\mu\nu\kappa}$ has been derived from the 3-form
$H^{(3)} = [(dx^\mu \wedge d x^\nu \wedge dx^\kappa)/ (3!)]\; H_{\mu\nu\kappa} 
\equiv d B^{(2)}$ which
is constructed with the help of the exterior derivative $d = dx^\mu \partial_\mu$ (with $d^2 = 0$)
and 2-form connection $B^{(2)} = [(dx^\mu \wedge dx^\nu) / ({2!})]\; B_{\mu\nu}$ that defines the tensor gauge field $B_{\mu\nu}$.

The six independent components  $B_{0i}$ and $B_{ij}$ of the antisymmetric tensor field $B_{\mu\nu}$ have the following expressions for their conjugate momenta:
$$
\begin{array}{lcl}
\Pi^{0i} = {\displaystyle \frac{\partial {\cal L}^{(0)}}{\partial_0 B_{0i}}} = H^{00i} = 0, 
\qquad
\Pi^{ij} = {\displaystyle \frac{\partial {\cal L}^{(0)}}{\partial_0 B_{ij}}} = H^{0ij}.
\end{array} \eqno(2.2)
$$
It is straightforward to note that $\Pi^{0i} \approx 0$ is the primary constraint on the theory in the language of Dirac's prescription for the classification scheme [1,2]. The secondary constraint $\partial_i \Pi^{ij} = \partial_j \Pi^{ij} \approx 0$ is nothing other than the equation of motion [14] with respect to the field component $B_{0i}$  
that is obtained due to the Euler-Lagrange equation of motion. This equation turns
out to be $\partial_\mu H^{\mu\nu\kappa} = 0$ from (2.1).

Taking into account the canonical definitions, it is clear that both the above constraints
(i.e. $\Pi^{0i} \approx 0$ and $\partial_i \Pi^{ij} \approx 0$) are first-class [14] in the language
of Dirac's prescription for the classification scheme. As a consequence, these constraints generate
a set of gauge symmetry transformations [1,2] for the Lagrangian density (2.1). The generator $G$ for these continuous symmetry transformations can be written in terms of the above constraints as
$$
\begin{array}{lcl}
G = {\displaystyle \int} \;d^3 x \;
\Bigl [ - (\partial_0 \Lambda_i - \partial_i \Lambda_0) \; \Pi^{0i}
+ (\partial_i \Pi^{ij})\; \Lambda_j - (\partial_j \Pi^{ij})\;\Lambda_i \Bigr ],
\end{array} \eqno(2.3)
$$
where the anti-symmetrization has been brought into the above expression 
because of the antisymmetric nature of 
$\Pi^{0i} = - \Pi^{i0}$ and $\Pi^{ij} = - \Pi^{ji}$. A partial integration leads to the following
simple expression for the above generator, namely; 
$$
\begin{array}{lcl}
G = {\displaystyle \int} \;d^3 x \;\Bigl [ - (\partial_0 \Lambda_i - \partial_i \Lambda_0) \; \Pi^{0i}
- \Pi^{ij}\; (\partial_i \Lambda_j - \partial_j \Lambda_i) \Bigr ].
\end{array} \eqno(2.4)
$$
Here $\Lambda_\mu$ (with $\Lambda_\mu = (\Lambda_0, \Lambda_i)$) are the infinitesimal gauge transformation parameters.

Exploiting the following equal-time canonical brackets (with $\hbar = c = 1$)
$$
\begin{array}{lcl}
&& [ B_{0i} ({\bf x}, t), \Pi^{0j} ({\bf y}, t) ] = i \;\delta^j_i \; \delta^{(3)} ({\bf x - y}), 
\nonumber\\
&& [ B_{ij} ({\bf x}, t), \Pi^{kl} ({\bf y}, t) ] = {\displaystyle \frac{i}{2}}\;
 (\delta^k_i \delta^l_j - \delta^k_j \delta^l_i) \; \delta^{(3)} ({\bf x - y}),
\end{array} \eqno(2.5)
$$
it is elementary to check that the following gauge transformations ($\delta_{gt}$) 
$$
\begin{array}{lcl}
\delta_{gt} B_{0i} = - \;(\partial_0 \Lambda_i - \partial_i \Lambda_0), \qquad
\delta_{gt} B_{ij} = - \; (\partial_i \Lambda_j - \partial_j \Lambda_i),
\end{array} \eqno(2.6)
$$
emerge\footnote{The minus signs
in the symmetry transformations have their origin in the minus signs present in the generator G
(cf. (2.4)). This choice has been taken into account for the later algebraic convenience.} 
from the generator $G$ if we exploit the following relationship (with $ {\bf x} \equiv x_i$ )
$$
\begin{array}{lcl}
 \delta_{gt} \Phi ({\bf x}, t) = - \;i \;[ \Phi ({\bf x}, t), G ], \qquad 
\Phi ({\bf x}, t) = B_{0i} ({\bf x}, t), B_{ij} ({\bf x}, t),
\end{array} \eqno(2.7)
$$
for the generic field $\Phi$ of the theory. The above transformations
are the symmetry transformations of the Lagrangian density (2.1) because 
$\delta_{gt} H_{\mu\nu\kappa} = 0$ under the transformations (2.6) which can be re-expressed,
in terms of the Greek indices, as
$\delta_{gt} B_{\mu\nu} = - (\partial_\mu \Lambda_\nu - \partial_\nu \Lambda_\mu)$.

Before we wrap up this section, a couple of key points are to be noted. First, there is a stage-one
reducibility in the theory because the following continuous transformation $\delta_f$ (with an
infinitesimal parameter $\omega$)
of the gauge transformation 
parameters $\Lambda_\mu (x)$\footnote{Here the gauge parameter function
has been treated like a field because this would be, later on, identified
with either an anti-ghost field or with a ghost field in the framework of BRST formalism.}
$$
\begin{array}{lcl}
\delta_f \Lambda_\mu = - \partial_\mu \omega \;\Rightarrow \;
\delta_f (\delta_{gt} B_{\mu\nu}) = 0 
\;\Rightarrow\;
\delta_f {\cal L}^{(0)} = 0,
\end{array} \eqno(2.8)
$$
leaves the Lagrangian density (2.1) invariant. However, this transformation is not generated
by the first-class constraints $\Pi^{0i} \approx 0, \partial_i \Pi^{ij} \approx 0$. Second, for
the consistent quantization of the theory under consideration, one has to impose the conditions
$\Pi^{0i} |phys> = 0$ and $(\partial_i \Pi^{ij}) |phys> = 0$ on the physical states 
(i.e. $|phys>$) of the quantum Hilbert space (i.e. Dirac's prescription). 
It is the BRST formalism which addresses both the above issues in a cogent manner. 
This is why, we shall discuss the same formalism in our next section.\\
 
\noindent
{\bf 3 BRST symmetry transformations: BRST charge}\\

\noindent
We begin with the Lagrangian density of the theory which is a generalization
of the Lagrangian density (2.1) such that it (i) respects the off-shell nilpotent
($ s_b^2 = 0$) BRST symmetry transformations $s_b$, and (ii) sheds light on the 
conceptual problems, encountered by the Lagrangian density (2.1),
that were pointed out at the fag end of the previous section.
Written in the explicit form, the above BRST invariant Lagrangian density reads, as [27]
$$
\begin{array}{lcl}
&&{\cal L}^{(b)}_B = {\displaystyle \frac{1}{6} H^{\mu\nu\kappa} H_{\mu\nu\kappa} 
+ B^\mu (\partial^\nu B_{\nu\mu}) +  \frac{1}{2} \Bigl (B \cdot B + \bar B \cdot \bar B \Bigr )
+ \partial_\mu \bar \beta \partial^\mu \beta - \frac{1}{2} \partial^\mu \phi
\partial_\mu \phi }\nonumber\\
&&+ \Bigl (\partial_\mu \bar C_\nu - \partial_\nu \bar C_\mu \Bigr ) \partial^\mu C^\nu
+ \Bigl (\partial \cdot C - \lambda \Bigr ) \rho + \Bigl (\partial \cdot \bar C + \rho 
\Bigr ) \lambda + L^\mu \Bigl (B_\mu - \bar B_\mu - \partial_\mu \phi \Bigr ),
\end{array} \eqno(3.1)
$$
where $B_\mu$ and $\bar B_\mu$ are the auxiliary fields, $\phi $ is a constrained
massless (i.e. $\Box \phi = 0$) scalar field, $L_\mu$ is a multiplier field, $\rho$ and
$\lambda$ are the fermionic ($\rho^2 = \lambda^2 = 0, \rho \lambda + \lambda \rho = 0$)
Lorentz scalar auxiliary ghost fields, $(\bar C_\mu) C_\mu$ are the fermionic 
($C_\mu^2 = \bar C_\mu^2 = 0, C_\mu \bar C_\nu + \bar C_\nu C_\mu = 0$, etc.) 
Lorentz vector (anti-)ghost fields with ghost number $(-1)1$, respectively, and $(\bar \beta)\beta$ are the bosonic (anti-)ghost fields with ghost number $(-2)2$, respectively.

The above Lagrangian density (3.1) respects\footnote{We adopt here the notations and conventions followed in [27]. In fact, the gauge parameters $\Lambda_\mu$ and $\omega$ of the transformations (2.6) and (2.8) have been replaced by the ghost fields $C_\mu$ and $\beta$, respectively, in the BRST transformations. The transformations $\delta_{gt}$ and $\delta_f$ {\it together}
have been replaced by the standard notation of the BRST transformation $s_b$ which is
found to be fermionic (i.e. $s_b^2 = 0$) in nature.} the following symmetry transformations
$$
\begin{array}{lcl}
&& s_b B_{\mu\nu} = - (\partial_\mu C_\nu - \partial_\nu C_\mu), \quad s_b C_\mu = - \partial_\mu \beta, \quad s_b \bar C_\mu = - B_\mu, \quad s_b L_\mu = - \partial_\mu \lambda, \nonumber\\
&& s_b \phi = \lambda, \qquad s_b \bar \beta = - \rho, \qquad s_b \bar B_\mu = - \partial_\mu \lambda,
\qquad s_b \bigl [\rho, \lambda, \beta, B_\mu, H_{\mu\nu\kappa} \bigr ] = 0,
\end{array} \eqno(3.2)
$$
because the Lagrangian density transforms to a total spacetime derivative as follows
$$
\begin{array}{lcl}
s_b {\cal L}^{(b)}_B = - \partial_\mu \Bigl [ (\partial^\mu C^\nu - \partial^\nu C^\mu) B_\nu +
\lambda B^\mu + \rho \partial^\mu \beta \Bigr ].
\end{array} \eqno(3.3)
$$
It will be noted that the BRST symmetry transformations in (3.2) are the generalization of the ``classical'' local gauge transformations (2.6) and (2.8) to the ``quantum'' level
as the latter are nilpotent (i.e. $s_b^2 = 0$) of order two
\footnote{One of the interpretations of this property 
is like setting $\hbar^2 \rightarrow 0$ in quantum mechanics.}. 
The Euler-Lagrange dynamical equations of motion, that emerge from the  Lagrangian density (3.1), are
as follows
$$
\begin{array}{lcl}
&& \Box \beta = \Box \bar \beta =  0, \qquad \Box \phi =  0, \qquad 
\Box \lambda = \Box \rho = 0, \qquad
\Box B_\mu = 0,  \nonumber\\
&& B_\mu - \bar B_\mu - \partial_\mu \phi = 0, \quad 
B_\mu + \bar B_\mu + \partial^\nu B_{\nu\mu} = 0,  \quad L_\mu = \bar B_\mu,
\quad \lambda = \frac{1}{2}\; (\partial \cdot C), \nonumber\\
&& \rho = - \frac{1}{2}\; (\partial \cdot \bar C), \quad \Box C_\mu = 
\frac{1}{2} \partial_\mu (\partial \cdot C) \equiv \partial_\mu \lambda,
\quad \Box \bar C_\mu = \frac{1}{2} \partial_\mu (\partial \cdot \bar C) 
\equiv - \partial_\mu \rho, \nonumber\\
&& \partial_\mu H^{\mu\nu\kappa} + \frac{1}{2}
(\partial^\nu B^\kappa - \partial^\kappa B^\nu) = 0,
\qquad  \partial \cdot B = \partial \cdot \bar B = \partial \cdot L = 0,
\end{array} \eqno(3.4)
$$
where $\Box = \partial_0^2 - \partial_i^2$ is the d'Alembertian operator and there is a complete
agreement between these dynamical equations of motion (cf. (3.4)) 
and the symmetry transformations (3.2). In fact, it is because of the 
requirement of the above ``agreement''
that equations $\Box \lambda = 0, \Box \rho = 0$ have been 
derived from the equations of motion $\Box \phi = 0, \Box \bar \beta = 0$.

It is clear that the auxiliary fields $B_\mu$ and $\bar B_\mu$ can be expressed as:
$$
\begin{array}{lcl}
B_\mu = - {\displaystyle \frac{1}{2}\; \Bigl [\partial^\nu B_{\nu\mu} - \partial_\mu \phi
\Bigr ], \qquad
\bar B_\mu = - \frac{1}{2}\; \Bigl [\partial^\nu B_{\nu\mu} + \partial_\mu \phi
\Bigr ]}.
\end{array} \eqno(3.5)
$$
In the above equalities too, there is a complete conformity between the symmetry transformations (3.2)
and the equations of motion (3.4). The Noether conserved current is
$$
\begin{array}{lcl}
J^\mu_{(b)} = (s_b \Phi_i) \; \Bigl ({\displaystyle \frac{\partial {\cal L}^{(b)}_B}{\partial_\mu \Phi_i}} \Bigr )
+ (\partial^\mu C^\nu - \partial^\nu C^\mu) B_\nu +
\lambda B^\mu + \rho \partial^\mu \beta,
\end{array} \eqno(3.6)
$$
where the generic field $\Phi_i = B_{\nu\kappa}, C_\nu, \bar C_\nu, \phi, \bar \beta$. The terms present in the square bracket of (3.3) have been added here for the computation of the precise expression for the Noether conserved current. The explicit form of this current is
$$
\begin{array}{lcl}
J^\mu_{(b)} &=& (\partial^\mu \bar C^\nu - \partial^\nu \bar C^\mu) \partial_\nu \beta
- \lambda (\partial^\mu \phi + L^\mu) - \rho \partial^\mu \beta \nonumber\\
&-& H^{\mu\nu\kappa}
(\partial_\nu C_\kappa - \partial_\kappa C_\nu) -  (\partial^\mu C^\nu - \partial^\nu C^\mu) B_\nu,
\end{array} \eqno(3.7)
$$
where $L_\mu = \bar B_\mu$  due to the equation of motion in (3.4).

It can be explicitly checked, using the equations of motion (3.4), that 
$$
\begin{array}{lcl}
\partial_\mu J^\mu_{(b)} = 0 \;\;\;\Rightarrow\;\;\; 
Q_{(b)} = {\displaystyle \int}\; d^3 x\; J^{(0)}_{(b)}, 
\end{array} \eqno(3.8)
$$
where $Q_{(b)}$ is the conserved BRST charge.
The explicit expression for $Q_{(b)}$ is:
$$
\begin{array}{lcl}
Q_{(b)} &=& {\displaystyle \int}\; d^3 x \; \Bigl [
(\partial^0 \bar C^i - \partial^i \bar C^0)\; \partial_i \beta
- \lambda \;(\partial^0 \phi + L^0) - \rho \;\partial^0 \beta \nonumber\\
&-& H^{0ij}\;
(\partial_i C_j - \partial_j C_i) -  (\partial^0 C^i - \partial^i C^0)\; B_i \Bigr ]. 
\end{array} \eqno(3.9)
$$
To check that the above conserved charge is the generator of the BRST symmetry transformations, it
is useful to express it in terms of the canonical momenta, that are derived from the Lagrangian density (3.1). These momenta are listed below
$$
\begin{array}{lcl}
&&\Pi^{ij} = H^{0ij}, \qquad \;\Pi^{0i} = B^i, \qquad \;\;\Pi_{(\phi)} = - (\partial^0 \phi + L^0),
\nonumber\\
&& \Pi_{(C_0)} = \rho, \qquad \;\;\Pi_{(\bar C_0)} = \lambda, \;\qquad \Pi^{i}_{({\bf C})}
= - (\partial^0 \bar C^i - \partial^i \bar C^0), \nonumber\\ && \Pi^{i}_{({\bf \bar C})}
= + (\partial^0 C^i - \partial^i  C^0), \qquad \Pi_{(\bar\beta)} = \partial^0 \beta,
\qquad \Pi_{(\beta)} = \partial^0 \bar \beta.
\end{array} \eqno(3.10)
$$
The expression for the BRST charge, in terms of (3.10), is
$$
\begin{array}{lcl}
Q_{(b)} &=& {\displaystyle \int}\; d^3 x \; \Bigl [
- \Pi^{i}_{({\bf C})}\; \partial_i \beta
+ \Pi_{(\bar C_0)}\;\Pi_{(\phi)} - \Pi_{(C_0)} \; \Pi_{(\bar\beta)} \nonumber\\
&-& \Pi^{ij}\; (\partial_i C_j - \partial_j C_i) - \Pi_{0i}\;\Pi^{i}_{({\bf \bar C})} \Bigr ].
\end{array} \eqno(3.11)
$$
A close look at the above equation clearly expresses the fact that it is the generalization
of the generator (2.4) where, besides momenta $\Pi^{ij}$ and $\Pi^{0i}$, the additional momenta
of the BRST invariant Lagrangian density (cf. (3.1)) are also present.

In addition to the equal-time canonical commutators (2.5), we have 
the following equal-time (anti)commutators (with $\hbar = c = 1$) in the theory, namely;
$$
\begin{array}{lcl}
&& \{ C_0 ({\bf x}, t), \rho ({\bf y}, t) \} = i \delta^{(3)} ({\bf x - y}), \qquad\;
\{ \bar C_0 ({\bf x}, t), \lambda ({\bf y}, t) \} = i \delta^{(3)} ({\bf x - y}), \nonumber\\
&& \{ C_i ({\bf x}, t), \Pi^j_{(\bf C)} ({\bf y}, t) \} 
= i \delta_i^j \delta^{(3)} ({\bf x - y}), \;
\{ \bar C_i ({\bf x}, t), \Pi^j_{(\bf \bar C)} ({\bf y}, t) \} 
= i \delta_i^j \delta^{(3)} ({\bf x - y}), \nonumber\\
&& [\beta ({\bf x}, t), \Pi_{(\beta)} ({\bf y}, t) ] 
= i \delta^{(3)} ({\bf x - y}), \qquad\;
[\bar\beta ({\bf x}, t), \Pi_{(\bar\beta)} ({\bf y}, t) ] 
= i \delta^{(3)} ({\bf x - y}), \nonumber\\
&& [ \phi ({\bf x}, t), \Pi_{(\phi)} ({\bf y}, t) ]  =  i \delta^{(3)} ({\bf x - y}), 
 \end{array} \eqno(3.12)
$$
and all the rest of the canonical (anti)commutators are zero. The above canonical brackets
demonstrate that the conserved charge $Q_{(b)}$ generates the nilpotent symmetry transformations (3.2). This statement can be expressed in terms the generic
field $\Phi_i$ of the theory and the conserved charge $Q_{(b)}$ as given below
$$
\begin{array}{lcl}
s_b \Phi_i = - i\; \Bigl [\;\Phi_i, Q_{(b)} \;\Bigr ]_{(\pm)}, \qquad \Phi_i = B_{\mu\nu}, C_\mu,
\bar C_\mu, B_\mu, \beta, \bar \beta, \phi, \lambda, \rho, 
\end{array} \eqno(3.13)
$$
where the $(\pm)$ signs on the square bracket stand for the (anti)commutator for the generic
field $\Phi_i$ being fermionic/bosonic in nature. It will be noted that the transformations for
the Lagrange multiplier field $L_\mu$ and the auxiliary field $\bar B_\mu$  are equal and same
because of the equation of motion $L_\mu = \bar B_\mu$. These transformations are derived from
the (anti-)BRST invariance of the constraint equation $B_\mu - \bar B_\mu - \partial_\mu \phi 
= 0 $ and the
anticommutativity of the (anti-)BRST symmetry transformations (see, Sec. 4 below, for the details).

The stage is now set to comment on the remarks made after equation (2.8). It can be
seen, from (3.13), that the transformations corresponding to the generalization of  (2.8)
(e.g. $s_b C_\mu = - \partial_\mu \beta$, etc.) are generated by the BRST charge. Since the
ghost fields are decoupled from the rest of the theory, the Hilbert space of states is the
direct product of the ghost states and the states of the rest of the fields of the theory.
The physical states of the theory are those which are annihilated by the BRST charge.
The following physicality condition with the conserved BRST charge, namely;
$$
\begin{array}{lcl}
Q_{(b)} |phys> = 0 \;\;\Rightarrow \;\;\Pi^{0i} |phys> = 0, \quad \partial_i \Pi^{ij} |phys> = 0,
\quad \Pi_{(\phi)} |phys> = 0, 
\end{array} \eqno(3.14)
$$
provides the reasons behind the consistent (i.e. Dirac's prescription)
as well as the covariant canonical quantization scheme (cf. (2.5) and (3.12))
for the theory. This is due to the fact that the following conditions 
are fulfilled, namely;

(i) the first-class constraints of the original theory 
annihilate the physical states of the quantum Hilbert space (i.e. $\Pi^{0i} |phys> = 0, 
\partial_i \Pi^{ij} |phys> = 0$), and 

(ii) none of the conjugate momenta (cf. (3.10)) are zero for the Lagrangian density (3.1)
(unlike (2.2) where $ \Pi^{0i} = 0$ for the Lagrangian density (2.1)). 

As a consequence, we do obtain the covariant canonical quantization of the theory. 
The well-defined canonical (anti)commutator brackets (2.5) and (3.12) exemplify it.

Using the equations of motion (3.4), it can be seen that 
$$
\begin{array}{lcl}
&&\Pi^{0i}\; |phys> = 0 \; \;\;\Rightarrow \; \;\;B^i |phys> = 0, \nonumber\\
&& \partial_j \Pi^{ij} \;|phys> = 0 \; \Rightarrow \; (\partial^0 B^i - \partial^i B^0) |phys> = 0, \nonumber\\
&& \Pi_{(\phi)} \;|phys> = 0  \;\;\Rightarrow \; \;(\partial^0 \phi + \bar B^0) |phys> = 0.
\end{array} \eqno(3.15)
$$ 
The last condition, ultimately, implies that $B^0 |phys> = 0$ due to the equations of motion (3.4). As a consequence, we have $B^\mu |phys> = 0$. Using (3.5), we conclude that the total gauge-fixing term (cf. Appendix A) annihilates
the physical states:
$$
\begin{array}{lcl}
 B^\mu |phys> = 0 \;\;\;\;\Rightarrow \;\;\; \;
(\partial_\nu B^{\nu\mu} - \partial^\mu \phi)\; |phys> = 0.
\end{array} \eqno(3.16)
$$ 
Exactly the same kind of condition is satisfied in the context of the application of the BRST
formalism to the 4D (non-)Abelian 1-form gauge theories.

A couple of comments are in order for the constraint conditions in (3.15) and (3.16). First and foremost,
it can be seen that these constraint conditions are BRST invariant because of the fact that
$s_b B_\mu = 0$ and $s_b (\partial^0 B^i - \partial^i B^0) = 0$. Second, the time derivatives of the 
constraint conditions $\Pi^{0i} |phys> \equiv B^i |phys> = 0$ and $\partial_i \Pi^{ij} |phys>
\equiv \partial_i H^{0ij} |phys>  = 0$ are also equal to zero and these are contained in the 
constraint conditions
listed in (3.15). For instance, it is trivial to check that $\partial_0 \partial_i H^{0ij} |phys> \equiv 
\partial_0 \partial_i \Pi^{ij} |phys> = 0$ because of the antisymmetry property of $H^{0ij}$ and 
symmetry property of
$\partial_0 \partial_i$. Furthermore, the condition $\partial_0 \Pi^{0i} |phys> \equiv \partial_0 B^i |phys> = 0$
is contained in the second entry of the equation (3.15). Thus, the physicality condition with
the conserved BRST charge (i.e. $Q_{(b)} |phys> = 0$)
ensures that the first-class constraint conditions (i.e. $\Pi^{0i} |phys> = 0, \partial_i \Pi^{ij} |phys> = 0$)
remain invariant w.r.t. the {\it time} evolution of the theory (Dirac's prescription) [1,2].\\

\noindent
{\bf 4 Anti-BRST symmetry transformations: anti-BRST  charge}\\

\noindent
The Lagrangian density ${\cal L}_B^{(ab)}$, that respects the off-shell nilpotent
($s_{ab}^2 = 0$) anti-BRST symmetry transformation $s_{ab}$, can be derived from
the BRST invariant Lagrangian density (3.1), by the substitution of the auxiliary
field $B^\mu = \bar B^\mu + \partial^\mu \phi$ in the gauge-fixing term
(i.e. $B^\mu (\partial^\nu B_{\nu\mu}$). The ensuing anti-BRST invariant Lagrangian
density is [27]
$$
\begin{array}{lcl}
&&{\cal L}^{(ab)}_B = {\displaystyle \frac{1}{6} H^{\mu\nu\kappa} H_{\mu\nu\kappa} 
+ \bar B^\mu (\partial^\nu B_{\nu\mu}) +  \frac{1}{2} \Bigl (B \cdot B + \bar B \cdot \bar B \Bigr )
+ \partial_\mu \bar \beta \partial^\mu \beta - \frac{1}{2} \partial^\mu \phi
\partial_\mu \phi }\nonumber\\
&&+ \Bigl (\partial_\mu \bar C_\nu - \partial_\nu \bar C_\mu \Bigr ) \partial^\mu C^\nu
+ \Bigl (\partial \cdot C - \lambda \Bigr ) \rho + \Bigl (\partial \cdot \bar C + \rho 
\Bigr ) \lambda + L^\mu \Bigl (B_\mu - \bar B_\mu - \partial_\mu \phi \Bigr ),
\end{array} \eqno(4.1)
$$
due to the fact that $\partial^\mu \phi (\partial^\nu B_{\nu\mu}) = \partial^\mu [\phi \partial^\nu
B_{\nu\mu}]$ is a total spacetime derivative.

The following off-shell nilpotent (i.e. $ s_{ab}^2 = 0$) anti-BRST transformations $s_{ab}$
$$
\begin{array}{lcl}
&& s_{ab} B_{\mu\nu} = - (\partial_\mu \bar C_\nu - \partial_\nu \bar C_\mu), \quad 
s_{ab} \bar C_\mu = - \partial_\mu \bar \beta, \quad s_{ab}  C_\mu = + \bar B_\mu, \nonumber\\
&& s_{ab} \phi = \rho, \;\qquad \; s_{ab}  \beta = - \lambda, \;\qquad \;s_{ab}  B_\mu = + \partial_\mu \rho,
\nonumber\\ &&s_{ab} L_\mu = - \partial_\mu \rho, \;\qquad\;\;
s_{ab} \bigl [\rho, \lambda, \bar \beta, \bar B_\mu, H_{\mu\nu\kappa} \bigr ] = 0,
\end{array} \eqno(4.2)
$$
are the symmetry transformations for the Lagrangian density (4.1) because the latter transforms
to a total spacetime derivative, under the former, in the following fashion:
$$
\begin{array}{lcl}
s_{ab} {\cal L}^{(ab)}_B = - \partial_\mu \Bigl [ (\partial^\mu \bar C^\nu - \partial^\nu \bar C^\mu) \bar B_\nu -
\rho \bar B^\mu + \lambda \partial^\mu \bar \beta \Bigr ].
\end{array} \eqno(4.3)
$$
The noteworthy points, at this stage, are 

(i) the transformations $s_{ab} B_\mu = \partial_\mu \rho$ and $s_{ab} L_\mu = - \partial_\mu \rho$
are consistent because the equation of motion is: $L_\mu + B_\mu = 0$. The other equations of motion that differ from (3.4) are: $\Box \bar B_\mu = 0$ and $\partial_\mu H^{\mu\nu\kappa} + \frac{1}{2}
(\partial^\nu \bar B^\kappa - \partial^\kappa \bar B^\nu) = 0$,

(ii) the BRST transformations (3.2) and the anti-BRST transformations (4.2) are absolutely anticommuting because $(s_b s_{ab} + s_{ab} s_b) \Phi_i = 0$ for the generic field $\Phi_i$
of the theory on the constraint surface, defined by the field equation
$B_\mu - \bar B_\mu - \partial_\mu \phi = 0$,

(iii) the above constraint surface emerges from both the Lagrangian densities (3.1) and (4.1)
and it is an (anti-)BRST invariant (i.e. $s_{(a)b} [B_\mu - \bar B_\mu - \partial_\mu \phi = 0] = 0$) quantity,

(iv) the Lagrangian densities (3.1) and (4.1) are the generalizations of the starting
Lagrangian density (2.1) and the symmetry transformations (2.6) and (2.8) have been
generalized to (3.2) (i.e. $\Lambda_\mu \to C_\mu, \omega \to \beta)$ as well as (4.2)
(i.e. $\Lambda_\mu \to \bar C_\mu, \omega \to \bar \beta)$.

The Noether conserved current $J^\mu_{(ab)}$ can be calculated for the above
anti-BRST symmetry transformations (4.2). The exact expression is 
$$
\begin{array}{lcl}
J^\mu_{(ab)} &=& (s_{ab} \Phi_i) \; 
\Bigl ({\displaystyle \frac{\partial {\cal L}^{(ab)}_B}{\partial_\mu \Phi_i}} \Bigr )
+ (\partial^\mu \bar C^\nu - \partial^\nu \bar C^\mu) \bar B_\nu -
\rho \bar B^\mu + \lambda \partial^\mu \bar \beta, \nonumber\\
&\equiv& - H^{\mu\nu\kappa} (\partial_\nu \bar C_\kappa - \partial_\kappa \bar C_\nu) 
- \rho (\partial^\mu  \phi + L^\mu)
- \lambda \partial^\mu \bar \beta \nonumber\\
&-& (\partial^\mu \bar C^\nu - \partial^\nu \bar C^\mu) \bar B_\nu
- (\partial^\mu C^\nu - \partial^\nu C^\mu) \partial_\nu \bar \beta,
\end{array} \eqno(4.4)
$$
where the generic field $\Phi_i = B_{\nu\kappa}, C_\nu, \bar C_\nu, \beta, \phi$. 
Exploiting the appropriate equations motion, it can be checked that $\partial_\mu J^\mu_{(ab)} = 0$. 
As a consequence, the conserved charge, that generates the nilpotent anti-BRST symmetry transformations (4.2), is 
$$
\begin{array}{lcl}
Q_{(ab)} = {\displaystyle \int} d^3 x J^{(0)}_{(ab)} &\equiv& - {\displaystyle \int}\; d^3 x \;
\Bigl [  (\partial^0  C^i - \partial^i  C^0)\; \partial_i \bar \beta
+ \rho \;(\partial^0 \phi + L^0) + \lambda \;\partial^0 \bar \beta \nonumber\\
&+& H^{0ij}\;
(\partial_i \bar C_j - \partial_j \bar C_i) +  (\partial^0 \bar C^i - \partial^i \bar C^0)\; \bar B_i \Bigr ]. 
\end{array} \eqno(4.5)
$$
The above conserved quantity can be expressed in terms of the canonical momenta derived from the Lagrangian density (4.1). It will be noted that the following modifications  
$$
\begin{array}{lcl}
\Pi_{(\phi)} = - (\partial^0 \phi + \bar B^0) \to \tilde \Pi_{(\phi)} = - (\partial^0 \phi - B^0),
\qquad \Pi^{0i} = B^i \to \tilde \Pi^{0i} = \bar B^i,
\end{array} \eqno(4.6)
$$
are required for the Lagrangian density (4.1) which differ from the canonical momenta (3.10)
that are derived from the BRST invariant Lagrangian density (3.1).

In terms of the appropriate canonical momenta (cf. (3.10) and (4.6)), we have the following expression
for the anti-BRST charge $Q_{(ab)}$, namely;
$$
\begin{array}{lcl}
Q_{(ab)} &=& {\displaystyle \int}\; d^3 x \; \Bigl [
- \Pi^{i}_{({\bf C})}\; \partial_i \bar \beta
+ \Pi_{(C_0)}\;\tilde \Pi_{(\phi)} - \Pi_{(\bar C_0)} \; \Pi_{(\beta)} \nonumber\\
&-& \Pi^{ij}\; (\partial_i \bar C_j - \partial_j \bar C_i) 
- \tilde \Pi_{0i}\;\Pi^{i}_{({\bf C})} \Bigr ].
\end{array} \eqno(4.7)
$$
It is straightforward to check that the above charge is the generator of the anti-BRST
symmetry transformations (4.2) as the analogue of the generic equation (3.13) is satisfied
with the replacements: $s_b \to s_{ab}, Q_{(b)} \to Q_{(ab)}$. For this verification, 
however,
the appropriate brackets from (2.5), (3.12) and the following brackets with (4.6), namely;
$$
\begin{array}{lcl}
&& [ \phi ({\bf x}, t), \tilde \Pi_{(\phi)} ({\bf y}, t) ]  
=  i \delta^{(3)} ({\bf x - y}),  \nonumber\\
&& [ B_{0i} ({\bf x}, t), \tilde \Pi^{0j} ({\bf y}, t) ]  
=  i \delta_i^j \delta^{(3)} ({\bf x - y}),
\end{array} \eqno(4.8)
$$
are to be used in the explicit computations. We have taken, in the above, $\hbar = c = 1$.

The physicality criterion (i.e. $  Q_{(ab)} |phys> = 0$)
with the conserved anti-BRST charge $Q_{(ab)}$ incorporates the first-class constraints of
the original theory. This statement can be mathematically expressed, in a succinct manner, as:
$$
\begin{array}{lcl}
Q_{(ab)} |phys> = 0 \;\;\Rightarrow \;\;\tilde \Pi^{0i} |phys> = 0, \quad 
\partial_i \Pi^{ij} |phys> = 0,
\quad \tilde \Pi_{(\phi)} |phys> = 0. 
\end{array} \eqno(4.9)
$$
The above restrictions, with the help of the appropriate equations of motion derived from the Lagrangian density (4.1), imply the following conditions on the physical states, namely;
$$
\begin{array}{lcl}
&& \tilde \Pi^{0i}\; |phys> = 0 \; \;\;\Rightarrow \;\;\;\bar B^i |phys> = 0, \nonumber\\
&& \partial_j \Pi^{ij} \;|phys> = 0 \; \Rightarrow \; (\partial^0 \bar B^i - \partial^i \bar B^0) |phys> = 0, \nonumber\\
&& \tilde \Pi_{(\phi)} \;|phys> = 0  \;\;\Rightarrow \; \;(\partial^0 \phi -  B^0) |phys> = 0
\;\;\Rightarrow \; \; \bar B^0 |phys> = 0.
\end{array} \eqno(4.10)
$$ 
The first and the last entries, in the above equation, imply that $\bar B^\mu |phys> = 0$. This,
in turn, leads to the result that the total gauge-fixing term also annihilates the physical state 
(i.e. $ (\partial^\nu B_{\nu\mu} + \partial_\mu \phi) |phys> = 0$) due to the equation\footnote{The total gauge-fixing term for the anti-BRST symmetry invariance can be taken to
be $(\partial_\nu B^{\nu\mu} + \partial^\mu \phi)$ as is evident from the Lagrangian density in (B.1) (see
Appendix B for details).} (3.5).

The above physicality condition (4.9), with the conserved  and nilpotent 
anti-BRST charge (i.e. $ Q_{(ab)} |phys> = 0$),
does ensure that

(i)  the first-class constraints of the original gauge theory, and

(ii) the time derivatives of the above constraints

annihilate the physical states of the theory. For the proof, the arguments go along the 
similar lines as the ones we have discussed for the physicality condition with the conserved 
BRST charge (cf. discussion after (3.16)). 
The constraint conditions in (4.10) are anti-BRST invariant because
$s_{ab} \bar B^\mu = 0$  and $s_{ab} (\partial^0 \bar B^i - \partial^i \bar B^0) = 0$.
Furthermore, it can be checked that $s_{ab} (\partial_\nu B^{\nu\mu} + \partial^\mu
\phi) = 0$ when we exploit the equations of motion for the auxiliary ghost field
$\rho = - \frac{1}{2} (\partial \cdot \bar C)$, Lorentz vector ghost field
$\Box \bar C_\mu = \frac{1}{2} \partial_\mu (\partial \cdot \bar C)$ and use the 
anti-BRST symmetry transformations (4.2).
However, under the BRST symmetry transformations (3.2), $s_b \bar B_\mu \neq 0$ and
$s_b (\partial_\nu B^{\nu\mu} + \partial^\mu \phi) \neq 0$ even if the on-shell conditions are used.
Only the constraint $(\partial^0 \bar B^i - \partial^i \bar B^0)$ of (4.10) remains invariant
under the BRST transformations (3.2) as well. Thus, in totality, the constraints (4.10) are not
BRST invariant. Similarly, the BRST invariant constraints of (3.15), in totality, are not 
found to remain invariant under the 
anti-BRST symmetry transformations (4.2). This observation establishes the physical independence of the anti-BRST
charge ($Q_{(ab)}$) {\it vis-{\`a}-vis} the BRST charge $Q_{(b)}$. 
Analogous result is {\it not} found in the context of the
4D Abelian 1-form gauge theory.

It would be very interesting to find out the BRST as well as anti-BRST invariant quantities from the above 
constraint analysis. This is important because $[s_b s_{ab} + s_{ab} s_b, s_b ] = 
[s_b s_{ab} + s_{ab} s_b, s_{ab} ] = 0$ shows that the anticommutator $\{ s_b, s_{ab} \} = (s_b + s_{ab})^2$ 
is a linearly independent combination w.r.t. nilpotent (i.e. $s_{(a)b}^2 = 0$) 
(anti-)BRST symmetry transformations $s_{(a)b}$.
A judicious guess, at this stage, is to look for a linear combination of $B_\mu$ and $\bar B_\mu$
(i.e. $B_\mu \pm \bar B_\mu$) with some other suitable fields for the (anti-)BRST invariant quantity. 
The above guidance comes from our discussions about the constraint equations (3.15) and (4.10)
where we have seen that $s_b B_\mu =0, s_{ab} \bar B_\mu = 0$ but $s_{ab} B_\mu \neq 0,  s_b \bar B_\mu \neq 0$.
It turns out that $s_{(a)b} [(B_\mu - \bar B_\mu - \partial_\mu \phi)] = 0$ and 
$s_{(a)b} [(B_\mu + \bar B_\mu + \partial^\nu B_{\nu\mu})] = 0$. For the latter, however, one has to
exploit the on-shell conditions $\Box C_\mu = \partial_\mu \lambda, \Box \bar C_\mu = - \partial_\mu \rho,
\lambda = \frac{1}{2} (\partial \cdot C), \rho = - \frac{1}{2} (\partial \cdot \bar C)$ from (3.4).
It is clear that, under the off-shell nilpotent (anti-)BRST symmetry transformations (3.2) and (4.2),
the combination $(B_\mu - \bar B_\mu - \partial_\mu \phi)$ remains invariant. This is why, it is this
condition that emerges from the superfield approach to BRST formalism [22] where the off-shell nilpotent
(anti-)BRST symmetry transformations have been considered.\\

\noindent
{\bf 5 Discrete and ghost symmetry transformations: ghost charge}\\

\noindent
It is interesting to note that the ghost part of the Lagrangian densities (3.1) and/or (4.1)
is endowed with a set of discrete symmetry transformations because it remains unchanged under the 
following transformations
$$
\begin{array}{lcl}
&& C_\mu \to \pm i \bar C_\mu, \quad \bar C_\mu \to \pm i  C_\mu, \quad \beta \to \pm i \bar \beta,
\quad  \bar \beta \to \mp i \beta, \quad \rho \to \mp i \lambda, \quad \lambda \to \mp i \rho, \nonumber\\
&& {\cal L}_{(g)} =  \partial_\mu \bar \beta \partial^\mu \beta 
+ \bigl (\partial_\mu \bar C_\nu - \partial_\nu \bar C_\mu \bigr ) \partial^\mu C^\nu
+ \bigl (\partial \cdot C - \lambda \bigr ) \rho + \bigl (\partial \cdot \bar C + \rho 
\bigr ) \lambda \to {\cal L}_{(g)},
\end{array} \eqno(5.1)
$$
where, as is evident, the Lagrangian density ${\cal L}_{(g)}$ is the ghost part of the 
(anti-)BRST invariant Lagrangian densities
(cf. (3.1), (4.1)) of the theory. In the above, signs at the top and at the bottom should
be taken {\it together} separately and independently. Thus, it is clear that the following
discrete transformations 
$$
\begin{array}{lcl}
&& C_\mu \to \pm i \bar C_\mu, \quad \bar C_\mu \to \pm i  C_\mu, \quad \beta \to \pm i \bar \beta,
\quad  \bar \beta \to \mp i \beta, \quad \rho \to \mp i \lambda, \quad \lambda \to \mp i \rho, \nonumber\\
&& B_{\mu\nu} \to B_{\mu\nu}, \qquad B_\mu \to B_\mu, \qquad \bar B_\mu \to \bar B_\mu, \qquad
\phi \to \phi, \qquad L_\mu \to L_\mu,
\end{array} \eqno(5.2)
$$ 
are the {\it symmetry} transformations for the total Lagrangian densities (3.1) and/or (4.1).

The discrete symmetry transformations for the ghost part of the Lagrangian densities provide
a clue to derive the anti-BRST symmetry transformations (4.2) from the BRST symmetry transformations
(3.2). For instance, it can be seen that the following transformations on the other bosonic fields
of the theory, namely;
 $$
\begin{array}{lcl}
B_{\mu\nu} \to \pm i B_{\mu\nu}, \qquad B_\mu \to \mp i \bar B_\mu, \qquad \bar B_\mu \to \pm i  B_\mu, \qquad
\phi \to \mp i \phi, \qquad L_\mu \to \mp i L_\mu,
\end{array} \eqno(5.3)
$$
together with the transformations (5.1), lead to the derivation of the anti-BRST symmetry
transformation $s_{ab}$ for any specific field of the theory from the corresponding
BRST symmetry transformation $s_b$ for exactly the same field. In other words, the BRST symmetry transformations $s_b$ go to the anti-BRST symmetry transformations $s_{ab}$ (i.e. $s_b \to s_{ab}$) under 
the combination of the discrete transformations (5.1) and (5.3).

As discussed above, in an exactly similar fashion, one can derive the BRST symmetry transformations
(3.2) from the anti-BRST symmetry transformations (4.2) if one exploits, in addition to the discrete ghost symmetry transformations (5.1), the following discrete
transformations on the gauge, scalar, multiplier and auxiliary fields, namely;  
$$
\begin{array}{lcl}
B_{\mu\nu} \to \pm i B_{\mu\nu}, \qquad B_\mu \to \pm i \bar B_\mu, \qquad \bar B_\mu \to \mp i  B_\mu, \qquad
\phi \to \mp i \phi, \qquad L_\mu \to \mp i L_\mu.
\end{array} \eqno(5.4)
$$
It should be noted that the transformations (5.4) are different from (5.3) because the auxiliary
fields $B_\mu$ and $\bar B_\mu$ transform in a different manner. In other words, the discrete 
transformations (5.1) and (5.4) blend together to produce the transformations $s_{ab} \to s_b$.

The ghost part of the Lagrangian density, given in (5.1), is endowed with a 
set of continuous global
symmetry transformations. The infinitesimal version (i.e. $\delta_g$) of these scale symmetry 
transformations for the above Lagrangian density (i.e. $\delta_g {\cal L}_g = 0$) 
are\footnote{It should be noted that the actual
transformations are: $ \beta \to e^{2 \Sigma} \beta, \bar \beta \to e^{- 2 \Sigma} \bar \beta,
\lambda \to e^{+ \Sigma} \lambda, \rho \to e^{- \Sigma} \rho$, etc.
Only the infinitesimal version of these transformations are quoted in (5.5).}
$$
\begin{array}{lcl}
&& \delta_g \beta = 2 \;\Sigma \;\beta, \qquad \delta_g \bar \beta = - 2 \;\Sigma \;\bar \beta,
\qquad  \delta_g C_\mu = + \;\Sigma \; C_\mu, \nonumber\\
&& \delta_g \bar C_\mu = - \;\Sigma \;\bar C_\mu, \qquad \delta_g \lambda = + \;\Sigma \;\lambda,
\qquad \delta_g \rho = - \;\Sigma \;\rho, 
\end{array} \eqno(5.5)
$$
where $\Sigma$ is a spacetime independent global parameter of the above scale transformations. The 
factors of $(\pm 2)$, appearing in the transformations for $\beta$ and $\bar\beta$,
correspond to the ghost numbers of these fields, respectively.  Similarly, the factors of $(\pm 1)$
in the transformations of $C_\mu$ and $\bar C_\mu$ fields are present because of the same reasons
as fermionic fields $(C_\mu)\bar C_\mu$ carry the ghost number equal to (+1) and (-1), respectively. 
It is interesting to point out that $\lambda = \frac{1}{2} (\partial \cdot C), \rho 
= -(1/2) (\partial \cdot \bar C)$ imply that, in the transformations of these fields 
(i.e. $\lambda, \rho)$), the factors of $(\pm 1)$ would appear which is the case in (5.5).
Thus, the auxiliary ghost fields $(\lambda, \rho)$ carry the ghost numbers $(+1, -1)$, respectively.

The continuous symmetry transformations (5.5) lead to the definition and derivation of the conserved Noether current. This conserved ghost current is
$$
\begin{array}{lcl}
J^\mu_{(g)} = 2 \beta \partial^\mu \bar \beta - 2 \bar \beta \partial^\mu \beta
+ (\partial^\mu C^\nu - \partial^\nu C^\mu) \bar C_\nu  
+ (\partial^\mu \bar C^\nu - \partial^\nu \bar C^\mu)  C_\nu + C^\mu \rho
- \bar C^\mu \lambda.
\end{array} \eqno(5.6)
$$ 
The above current is conserved as can be readily checked, using the equations of motion (3.4),
that $\partial_\mu J^\mu_{(g)} = 0$. This result implies that there is a conserved ghost charge
given by 
$$
\begin{array}{lcl}
Q_{(g)} =
{\displaystyle \int } d^3 x  \Bigl [ 2 \beta \Pi_{(\beta)} - 2 \bar \beta \Pi_{(\bar\beta)}
+ C_i \Pi^i_{({\bf C})} 
- \bar C_i \Pi^{i}_{({\bf \bar C})}  + C^0  \Pi_{(C_0)} 
- \bar C^0 \Pi_{(\bar C_0)} \Bigr ].
\end{array} \eqno(5.7)
$$
The above charge is the generator of the transformations (5.5), as can be verified
from the generic equation (3.13), by the replacements: $s_b \to \delta_g, Q_{(b)} \to Q_{(g)}$.
Furthermore, the canonical (anti)commutator brackets of (3.12) have to be exploited for the explicit proof of the above statement when the infinitesimal version (i.e. $\delta_g$) of the
scale transformations (cf. (5.5)), corresponding to the (fermionic)bosonic ghost fields, are to be derived.\\

\noindent
{\bf 6 BRST algebra: cohomological aspects}\\

\noindent
The conserved charges (which are the generators of the nilpotent (anti-)BRST symmetry 
transformations as well as the ghost scale transformations) obey the standard BRST algebra.
A similar kind of algebra is also obeyed by the
transformation operators $s_{(a)b}$ and $\delta_g$. For instance, the following algebra 
$$
\begin{array}{lcl}
&&  \{ s_b, s_{ab} \} \equiv s_b s_{ab} + s_{ab} s_b = 0, \quad s_b^2 \equiv \frac{1}{2} 
\{ s_b, s_b \} = 0, 
\quad s_{ab}^2 \equiv \frac{1}{2} \{ s_{ab}, s_{ab} \} = 0, \nonumber\\
&& [\delta_g, s_b ] \equiv \delta_g s_b - s_b \delta_g = + s_b, \qquad 
[\delta_g, s_{ab} ] \equiv \delta_g s_{ab} - s_{ab} \delta_g = - s_{ab},
\end{array} \eqno(6.1)
$$
ensues from the operation of the transformations $s_{(a)b}$ and $\delta_g$ on any arbitrary
generic field $\Phi_i = B_{\mu\nu}, B_\mu, \bar B_\mu, \phi, C_\mu, \bar C_\mu, \beta, \bar\beta,
\lambda, \rho$ of the Lagrangian densities (3.1) and (4.1).

The anticommutativity property
(i.e. $s_b s_{ab} + s_{ab} s_b = 0$) of the nilpotent
(anti-)BRST transformations $s_{(a)b}$ is valid only on 
a constrained surface, defined by the field equation $B_\mu - \bar B_\mu - \partial_\mu \phi = 0$,
on the 4D spacetime manifold. This property is similar to the 4D non-Abelian 1-form gauge theory
where the anticommutativity of the off-shell
nilpotent (anti-)BRST symmetry transformations $s_{(a)b}$ is satisfied due to the Curci-Ferrari restriction [27]. Furthermore, it will be noted 
that all the numerical factors (i.e. the ghost numbers),
that are given in the transformations (5.5), are very essential in the proof of
the algebraic relations: $[\delta_g, s_b ] = \;+ \;s_b$ and $[\delta_g, s_{ab} ] = \;- \;s_{ab}$.

The algebra (6.1) can be replicated in the language of the conserved (anti-)BRST and
ghost charges by exploiting the canonical (anti)commutators given in (2.5), (3.12)
and (4.8). This is logical because the above conserved charges are the generators of the 
infinitesimal transformations $s_{(a)b}$ and $\delta_g$. These charges satisfy the following BRST algebra  
$$
\begin{array}{lcl}
&&  \{ Q_{(b)}, Q_{(ab)} \} = 0, 
\quad Q_{(b)}^2 \equiv \frac{1}{2} 
\{ Q_{(b)}, Q_{(b)} \} = 0, 
\quad Q_{ab}^2 \equiv \frac{1}{2} \{ Q_{(ab)}, Q_{(ab)} \} = 0, \nonumber\\
&& i \;[ Q_{(g)}, Q_{(b)} ] = \;+ \;Q_{(b)}, \qquad 
i\;[ Q_{(g)}, Q_{(ab)} ]  = \;- \;Q_{(ab)}.
\end{array} \eqno(6.2)
$$
With the help of the above algebra, it can be seen that an arbitrary state with
the ghost number $n$ (i.e. $i Q_{(g)} |\psi>_n = n \; |\psi>_n$) imply the following
relationships
$$
\begin{array}{lcl}
i\; Q_{(g)} \; Q_{(b)} \; |\psi>_n = (n + 1)\; Q_{(b)}\; |\psi>_n, \quad
i\; Q_{(g)} \; Q_{(ab)} \; |\psi>_n = (n - 1)\; Q_{(ab)}\; |\psi>_n.
\end{array} \eqno(6.3)
$$
The above equation shows that the ghost numbers of the states $Q_{(b)} |\psi>_n$
and $Q_{(ab)} |\psi>_n$ are $(n + 1)$ and $(n - 1)$, respectively. In other words,
the BRST charge $Q_{(b)}$ increases the ghost number of a state by one. On the contrary, the
ghost number of a state is decreased by one when the operation of the anti-BRST charge 
$Q_{(ab)}$ is performed on it.

The above kind of properties are also found in the case of differential geometry
when the de Rham cohomological operators $d = dx^\mu \partial_\mu$ (with  $d^2 = 0$) 
and $\delta = - * d *$ (with $ \delta^2 = 0$) operate on the differential forms. Here
$d$ and $\delta$ are the exterior and co-exterior derivatives and $*$ is the
Hodge duality operation. It turns out that the operation of (i) $d$ increases the
degree of a form by one, and (ii) $\delta$ decreases the degree of a form by one.
However, the cohomological operators $d$ and $\delta$ {\it cannot} be identified with
the BRST and anti-BRST charges  because the latter anticommute with each-other (cf.
(6.2)) but the former don't. In fact, the anticommutator of $d$ and $\delta$ define
the Laplacian operator of the de Rham cohomology. An attempt has been made, in our 
earlier work [18], to establish a connection between the de Rham cohomological
operators and the conserved charges that generate the symmetry transformations
for a given Lagrangian density of the Abelian 2-form gauge theory. However, our 
current Lagrangian density is different from the one discussed in [18].

There is a simpler way to derive the BRST algebra (6.2) by exploiting the infinitesimal
BRST symmetry transformations (3.2), anti-BRST symmetry transformations (4.2) and 
the ghost transformations (5.5). This can be achieved by taking the help of the conserved 
BRST, anti-BRST and ghost charges as the generators (cf. (3.13))
of the above infinitesimal transformations.
Mathematically, the above claims are expressed as follows
$$
\begin{array}{lcl}
&& s_b\; Q_{(g)} = - i [Q_{(g)}, Q_{(b)}] = - Q_{(b)}, \qquad
s_{ab}\; Q_{(g)} = - i [Q_{(g)}, Q_{(ab)}] = + Q_{(ab)}, \nonumber\\
&& s_b Q_{(b)}
= - i \{ Q_{(b)}, Q_{(b)} \} = 0, \qquad
\delta_g\; Q_{(b)} = - i [Q_{(b)}, Q_{(g)}] = + Q_{(b)}, \nonumber\\
&& \delta_{g}\; Q_{(ab)} = - i [Q_{(ab)}, Q_{(g)}] = - Q_{(ab)}, \qquad s_{ab} Q_{(ab)}
= - i \{ Q_{(ab)}, Q_{(ab)} \} = 0, \nonumber\\
&& s_b Q_{(ab)} = - i \{ Q_{(ab)}, Q_{(b)} \} = 0, \qquad
s_{ab} Q_{(b)} = - i \{ Q_{(b)}, Q_{(ab)} \} = 0.
\end{array} \eqno(6.4)
$$
In the above computation, one has to take into account the expressions for the
BRST charge (3.9), anti-BRST charge (4.7) and the ghost charge (5.7). It is worthwhile 
to mention that the equations (6.1), (6.2) and (6.4) represent the same algebraic 
structure in somewhat different looking garbs where, primarily,
 the nilpotency and anticommutativity properties
of the (anti-)BRST charges and their corresponding transformations are intertwined.

Before wrapping up this section, we would like to make some useful remarks. It is very easy to
compute the brackets $- i [ Q_{[(a)b]}, Q_{(g)} ] = \delta_g Q_{[(a)b]}$ from the r.h.s.
(i.e. $\delta_g Q_{[(a)b]}$)
to show that $i [Q_{(g)}, Q_{[(a)b]}] = \mp Q_{[(a)b]}$ because the ghost transformations
(5.5) are very simple. It is also to be noted that all the numerical factors 
(i.e. $\pm 1, \pm 2$, etc.), present
in (5.5), turn out to be very essential in this proof. However, the computation of the same
brackets from $i [ Q_{(g)}, Q_{[(a)b]} ] = s_{[(a)b]} Q_{(g)}$ requires use of the equations
of motion listed in (3.4) as well as the ones derived from the Lagrangian density (4.1).
In particular, the calculation  of the transformations $s_b Q_{(ab)} \equiv s_{ab} Q_{(b)}$,
that lead to the computation of the bracket $\{ Q_{(b)}, Q_{(ab)} \}$, always need the help
from the equations of motion $B_\mu - \bar B_\mu - \partial_\mu \phi = 0$ and
$B_\mu + \bar B_\mu + \partial^\nu B_{\nu\mu} = 0$ as well as their off-shoots that are mentioned
in the equation (3.5). The proof of the nilpotency (i.e. $Q_{[(a)b]}^2 = 0$) of the charges $Q_{[(a)b]}$, from the transformations $s_{(a)b} Q_{[(a)b]}$, is quite straightforward. It leads to
$\{Q_{(b)}, Q_{(b)} \} = 0$ and $\{Q_{(ab)}, Q_{(ab)} \} = 0$. \\

\noindent
{\bf 7 Conclusions}\\

\noindent
One of the key points of our present investigation is the derivation of the
off-shell nilpotent (anti-)BRST charges and the ghost charge which obey the standard
BRST algebra and are the generators of the infinitesimal transformations
(3.2), (4.2) and (5.5). The interesting results of our present endeavor are

(i) the derivation of the (anti-)BRST invariant Curci-Ferrari type 
constraint field condition $B_\mu - \bar B_\mu - \partial_\mu \phi = 0$ from
the physicality\footnote{The common equations of motion, emerging from the Lagrangian
densities (3.1) and (4.1), are always (anti-)BRST invariant quantities. However, the
condition $B_\mu - \bar B_\mu - \partial_\mu \phi = 0$ is special because of our constraint 
analysis performed in (3.15) and (4.10) (see, e.g., the fag end of Sec. 4 for more discussions).} 
criteria $Q_{[(a)b]} |phys> = 0$ (cf. Sec. 4), 

(ii) the proof of the independent identity of the anti-BRST charge, and

(iii) the constraint analysis that enables us to achieve a consistent quantization.

It is worth noting that, the very same condition of (i), was derived 
in our earlier work [22], from the geometrical superfield approach to BRST
formalism where the nilpotency and anticommutativity of the (anti-)BRST
symmetry transformations are ensured because of their intimate connections
with the translational generators along the Grassmannian directions of
the (4, 2)-dimensional supermanifold on which the basic fields of the ordinary gauge 
theory are generalized to the corresponding superfields (see, e.g. [22] for details).

It may be emphasized that, in our present work, the anticommutativity
property $(s_b s_{ab} + s_{ab} s_b) \;B_{\mu\nu} = 0$ is satisfied only
due to the constrained field equation $B_\mu - \bar B_\mu - \partial_\mu \phi = 0$.
This relationship, within the framework of our present BRST formalism,
emerges because the conserved 
BRST and anti-BRST charges play their {\it independent} roles.
Analogous situation arises for the requirement of the 
anticommutativity property of the (anti-)BRST symmetry transformations
in the context of the non-Abelian 1-form ($A^{(1)} = dx^\mu A_\mu$)
gauge theory where the Curci-Ferrari restriction is invoked (see, e.g. [23]).
In yet another piece of our earlier work [27], the above constrained field equation
$B_\mu - \bar B_\mu - \partial_\mu \phi = 0$ has been shown to owe its origin
to the geometrical objects called gerbes which have been claimed to 
appear {\it always} 
in the context of 4D higher p-form ($p \geq 2$) Abelian gauge theories [27].

There are a few key differences between the Curci-Ferrari condition [23]
of the 4D {\it non-Abelian} 1-form gauge theory and the {\it one} encountered in
our present discussion of the 4D {\it Abelian}  2-form gauge theory. First,
the Curci-Ferrari condition involves the auxiliary fields as well as the
fermionic (anti-)ghost fields of the non-Abelian 1-form gauge theory but
our condition $B_\mu - \bar B_\mu - \partial_\mu \phi = 0$ comprises
of the auxiliary fields and a massless ($ \Box \phi = 0$) scalar field
$\phi$. Second, the auxiliary fields of the non-Abelian 1-form gauge theory are 
Lorentz scalars but, in our case of 2-form Abelian theory, the auxiliary
fields are Lorentz vectors. Finally, there is no need of incorporating the Curci-Ferrari
restriction, through a multiplier field, in the Lagrangian density of the
non-Abelian 1-form gauge theory. However, in our Abelian 2-form gauge theory,
it is very essential to include this restriction in the Lagrangian
density of the theory (cf. (3.1) and (4.1)) for the (anti-)BRST symmetry
invariance.

It is well known that there exists an off-shell as well as on-shell nilpotent BRST symmetry
transformation for a specific Lagrangian density of the 4D non-Abelian gauge theory
(see, e.g. [29,30] for details). 
However, there is {\it no} off-shell and/or on-shell nilpotent anti-BRST symmetry transformations
for the same specific Lagrangian density of the theory. This non-existence
has its origin\footnote{Another
explanation for the non-existence of the on-shell nilpotent anti-BRST symmetry transformations
for the non-Abelian 1-form gauge theory has been provided in the framework of the
superfield approach to BRST formalism (see, e.g. [29,30] for details).} 
in the existence of the Curci-Ferrari 
condition which allows only the off-shell nilpotent and anticommuting (anti-)BRST
symmetry transformations to exist. However, a specific set of Lagrangian densities of 
the Abelian 2-form gauge theory does allow {\it only} the 
presence of the on-shell as well as off-shell nilpotent BRST symmetries.
Similarly, another specific set of Lagrangian densities, for the present 2-form theory,
respects the off-shell
as well as on-shell nilpotent anti-BRST
symmetry transformations
(where there is no constrained field equation like 
$B_\mu - \bar B_\mu - \partial_\mu \phi = 0$). We have captured these results
in our Appendices A and B in a very concise manner.

It is clear, from Sec. 6, that the cohomological aspects of our present 4D
free Abelian 2-form gauge theory (described by the Lagrangian densities (3.1)
and (4.1)) are not yet complete. It would be a very nice endeavor to modify
the above Lagrangian densities so as to find out the dual-BRST and
anti-dual-BRST symmetries of the theory. This would enable us to find out the analogue
of the co-exterior derivative and Laplacian operators of the de Rham cohomology.
This is a challenging problem for us for our future
investigations because these results
would render the theory to become a model of Hodge theory. To capture the above 
symmetries in the language of the superfield approach
to BRST formalism is yet another direction for further research. It would be interesting to
apply the BRST approach to a
field theoretical model [31] where the 2-form gauge potential appears
in an intriguing fashion. Furthermore, it would be very interesting endeavor
to extend our earlier work [27]
to establish a connection between the Curci-Ferrari type of restriction and geometrical
objects, called gerbes, in the context of non-Abelian 2-form gauge theory. The latter are currently
one of the very active areas of research [32-34] in theoretical high energy physics. 
We shall report about all these results in our future publications [35].
 
\begin{center}
{\bf Appendix A}
\end{center}
 
\noindent
We discuss here the off-shell as well as on-shell nilpotent BRST symmetry
transformations for the present 4D Abelian 2-form gauge theory. The following Lagrangian density 
$$
\begin{array}{lcl}
{\cal L}^{(0)}_B &=& {\displaystyle \frac{1}{6} H^{\mu\nu\kappa} H_{\mu\nu\kappa} 
+  B^\mu (\partial^\nu B_{\nu\mu} - \partial_\mu \phi) +  \frac{1}{2} B \cdot B 
+ \partial_\mu \bar \beta \partial^\mu \beta  }\nonumber\\
&+& \Bigl (\partial_\mu \bar C_\nu - \partial_\nu \bar C_\mu \Bigr ) \partial^\mu C^\nu
+ \Bigl (\partial \cdot C - \lambda \Bigr ) \rho + \Bigl (\partial \cdot \bar C + \rho 
\Bigr ) \lambda,
\end{array} \eqno(A.1)
$$ 
is the limiting case of the Lagrangian densities in (3.1) and (4.1) where
there is no restriction like $B_\mu - \bar B_\mu - \partial_\mu \phi = 0$. The above
Lagrangian density is endowed with the following off-shell nilpotent ($s_b^2 = 0$)
BRST symmetry transformations $s_b$, namely; 
$$
\begin{array}{lcl}
&& s_b B_{\mu\nu} = - (\partial_\mu C_\nu - \partial_\nu C_\mu), \quad s_b C_\mu = - \partial_\mu \beta, \quad s_b \bar C_\mu = - B_\mu, \nonumber\\
&& s_b \phi = \lambda, \qquad \;s_b \bar \beta = - \rho, 
\qquad \;s_b \bigl [\rho, \lambda, \beta, B_\mu, H_{\mu\nu\kappa} \bigr ] = 0,
\end{array} \eqno(A.2)
$$
because the above Lagrangian density is quasi-invariant under (A.2) as
it transforms to a total spacetime derivative (i.e.
$s_b {\cal L}^{(0)}_B = - \partial_\mu [ (\partial^\mu C^\nu - \partial^\nu C^\mu) B_\nu +
\lambda B^\mu + \rho \partial^\mu \beta ]$).

It can be checked that the following equations of motion
$$
\begin{array}{lcl}
B_\mu = - (\partial^\nu B_{\nu\mu} - \partial_\mu \phi), \qquad \lambda = 
{\displaystyle \frac{1}{2}}\; (\partial \cdot C), \qquad
\rho = - 
{\displaystyle \frac{1}{2}}\; (\partial \cdot \bar C), 
\end{array} \eqno(A.3)
$$
allow us to derive an on-shell (i.e. $ \Box \phi = 0, \Box \beta = 0,
\Box C_\mu = (1/2) \;\partial_\mu (\partial \cdot C)$)
nilpotent (i.e. $\tilde s_b^2 = 0$) BRST symmetry transformations $\tilde s_b$ from (A.2) as
given below
$$
\begin{array}{lcl}
&& \tilde s_b B_{\mu\nu} = - (\partial_\mu C_\nu - \partial_\nu C_\mu), \quad 
\tilde s_b C_\mu = - \partial_\mu \beta, \quad 
\tilde s_b \bar C_\mu = + (\partial^\nu  B_{\nu\mu}
- \partial_\mu \phi), \nonumber\\
&& \tilde s_b \phi = {\displaystyle \frac{1}{2}} (\partial \cdot C), \qquad \;
\tilde s_b \bar \beta = 
{\displaystyle \frac{1}{2}} (\partial \cdot \bar C), 
\qquad \;\tilde s_b [\beta,  H_{\mu\nu\kappa}]  = 0.
\end{array} \eqno(A.4)
$$
These on-shell nilpotent transformations
are the symmetry transformations for the following Lagrangian density which is derived
from (A.1) due to (A.3), namely;
$$
\begin{array}{lcl}
 \tilde {\cal L}^{(0)}_B &=& {\displaystyle \frac{1}{6} H^{\mu\nu\kappa} H_{\mu\nu\kappa} 
- {\displaystyle \frac{1}{2}} (\partial^\nu B_{\nu\mu} - \partial_\mu \phi)^2 
+ \partial_\mu \bar \beta \partial^\mu \beta  } \nonumber\\ 
&+& \Bigl (\partial_\mu \bar C_\nu - \partial_\nu \bar C_\mu \Bigr ) \partial^\mu C^\nu
+ {\displaystyle \frac{1}{2}} \; \Bigl (\partial \cdot \bar C \Bigr ) 
\Bigl (\partial \cdot C \Bigr ).
\end{array} \eqno(A.5)
$$ 
The above Lagrangian density transforms to a total spacetime derivative under (A.4) . It is worthwhile to mention that the Lagrangian densities in (A.1) and (A.5) are {\it not} endowed with the anticommuting anti-BRST symmetry transformations.

Some attempts have been made earlier 
(see., e.g. [18] for details) to
find out the anti-BRST symmetry transformations for the above Lagrangian densities but
the nilpotent (anti-)BRST symmetry transformations are found {\it not} to be {\it absolutely} anticommuting
in nature. They are, instead, found to be off-shell nilpotent but anticommuting {\it only} up to a 
U(1) vector Abelian 
gauge transformation [18-20].\\

\begin{center}
{\bf Appendix B}
\end{center}

\noindent
We begin with the following Lagrangian density of the 4D Abelian 2-form
gauge theory: 
$$
\begin{array}{lcl}
{\cal L}^{(0)}_{\bar B} &=& {\displaystyle \frac{1}{6} H^{\mu\nu\kappa} H_{\mu\nu\kappa} 
+  \bar B^\mu (\partial^\nu B_{\nu\mu} + \partial_\mu \phi) +  \frac{1}{2} \bar B \cdot \bar B 
+ \partial_\mu \bar \beta \partial^\mu \beta  }\nonumber\\
&+& \Bigl (\partial_\mu \bar C_\nu - \partial_\nu \bar C_\mu \Bigr ) \partial^\mu C^\nu
+ \Bigl (\partial \cdot C - \lambda \Bigr ) \rho + \Bigl (\partial \cdot \bar C + \rho 
\Bigr ) \lambda.
\end{array} \eqno(B.1)
$$ 
The above Lagrangian density is the limiting case of the the Lagrangian densities (3.1)
and (4.1) where there is no restriction like $B_\mu - \bar B_\mu - \partial_\mu \phi = 0$.
Under the following off-shell nilpotent ($s_{ab}^2 = 0$) 
anti-BRST symmetry transformations $s_{ab}$
$$
\begin{array}{lcl}
&& s_{ab} B_{\mu\nu} = - (\partial_\mu \bar C_\nu - \partial_\nu \bar C_\mu), 
\quad s_{ab} \bar C_\mu = - \partial_\mu \bar \beta, \quad s_{ab}  C_\mu = + \bar B_\mu, \nonumber\\
&& s_{ab} \phi = \rho, \qquad \;s_{ab}  \beta = - \lambda, 
\qquad \;s_{ab} \bigl [\rho, \lambda, \bar \beta, \bar B_\mu, H_{\mu\nu\kappa} \bigr ] = 0,
\end{array} \eqno(B.2)
$$
the Lagrangian density (B.1) changes as:  
$s_{ab} {\cal L}^{(0)}_{\bar B} = - \partial_\mu [ (\partial^\mu \bar C^\nu 
- \partial^\nu \bar C^\mu) \bar B_\nu -
\rho \bar B^\mu + \lambda \partial^\mu \bar \beta ]$.

The following Euler-Lagrange equations of motion
$$
\begin{array}{lcl}
\bar B_\mu = - (\partial^\nu B_{\nu\mu} + \partial_\mu \phi), \qquad \lambda = 
{\displaystyle \frac{1}{2}}\; (\partial \cdot C), \qquad
\rho = - 
{\displaystyle \frac{1}{2}}\; (\partial \cdot \bar C), 
\end{array} \eqno(B.3)
$$
enable us to derive the on-shell (i.e. $ \Box \phi = 0, \Box \bar \beta = 0,
\Box \bar C_\mu = \frac{1}{2} \;\partial_\mu (\partial \cdot \bar C)$) nilpotent 
($\tilde s_{ab}^2 = 0$) anti-BRST symmetry transformations $\tilde s_{ab}$ from
the off-shell nilpotent transformations (B.2). The ensuing transformations are listed as follows
$$
\begin{array}{lcl}
&& \tilde s_{ab} B_{\mu\nu} = - (\partial_\mu \bar C_\nu - \partial_\nu \bar C_\mu), \quad 
\tilde s_{ab} \bar  C_\mu = - \partial_\mu \bar \beta, \quad 
\tilde s_{ab}  C_\mu = - (\partial^\nu  B_{\nu\mu}
+ \partial_\mu \phi), \nonumber\\
&& \tilde s_{ab} \phi =  - {\displaystyle \frac{1}{2}} (\partial \cdot \bar C), \qquad \;
\tilde s_{ab}  \beta = -
{\displaystyle \frac{1}{2}} (\partial \cdot  C), 
\qquad \;\tilde s_{ab} [\bar \beta,  H_{\mu\nu\kappa}]  = 0.
\end{array} \eqno(B.4)
$$
The above on-shell nilpotent 
transformations are the {\it symmetry} transformations for the following Lagrangian density
$$
\begin{array}{lcl}
 \tilde {\cal L}^{(0)}_{\bar B} &=& {\displaystyle \frac{1}{6} H^{\mu\nu\kappa} H_{\mu\nu\kappa} 
- {\displaystyle \frac{1}{2}} (\partial^\nu B_{\nu\mu} + \partial_\mu \phi)^2 
+ \partial_\mu \bar \beta \partial^\mu \beta  } \nonumber\\ 
&+& \Bigl (\partial_\mu \bar C_\nu - \partial_\nu \bar C_\mu \Bigr ) \partial^\mu C^\nu
+ {\displaystyle \frac{1}{2}} \; \Bigl (\partial \cdot \bar C \Bigr ) 
\Bigl (\partial \cdot C \Bigr ),
\end{array} \eqno(B.5)
$$ 
which is derived from the Lagrangian density (B.1) by using the equations of motion (B.3).

It can be readily seen that the off-shell as well as on-shell
nilpotent BRST and anti-BRST charges can
be calculated from the Noether's theorem for the transformations (A.2), (B.2), (A.4) and (B.4). 
The ghost symmetry transformations would be the same as
discussed in Sec. 5. The key point to be noted is that an absolutely 
anticommuting anti-BRST symmetry transformation (of the on-shell or off-shell variety)
does not exist for the Lagrangian densities (A.1) and (A.5) corresponding to the
symmetry transformations (A.2) and (A.4), respectively. In an exactly similar fashion, the absolutely
anticommuting BRST symmetry transformations of any variety, for the anti-BRST invariant
Lagrangian densities (B.1) and (B.5), do not exist. The off-shell nilpotent and anticommuting
(anti-)BRST symmetry transformations exist only for the Lagrangian densities (3.1) and (4.1).\\

\noindent
{\bf Acknowledgements:}
The present investigation has been carried out under a project entitled
``BRST symmetries and supersymmetries''.
Financial support from the Department of Science and Technology, 
Government of India, is gratefully acknowledged. It is a pleasure to thank
Saurabh Gupta, B. P. Mandal and Sanjay Siwach for discussions.

\end{document}